# Situational Preparedness Dynamics for Sequential Tropical Cyclone Hazards

Tianle Duan[1,*], Qingchun Li[1], Fengxiu Zhang[2], Dazhi Xi[3], Ning Lin[4]

[1]School of Construction Management Technology, Purdue University, West Lafayette, IN, USA

[2]Schar School of Policy and Government, George Mason University, Mason Square, VA, USA

[3]Department of Earth Sciences, The University of Hong Kong, Hong Kong SAR, China

[4]Department of Civil and Environmental Engineering, Princeton University, Princeton, NJ, USA

*Corresponding email: duan114@purdue.edu

## Abstract

Sequential tropical cyclone hazards—two tropical cyclones (TCs) making landfall in the same region within a short time—are becoming increasingly likely. This study investigates situational preparedness dynamics for six sequential TC events that affected seven states in the United States from 2020 to 2024. We find a combined effect of forecast wind speed and landfall sequence of a TC. Stronger forecast wind is always associated with higher preparedness levels. People tend to show a higher preparedness level for the second TC but are more sensitive to the increasing forecast wind speed of the first TC. We also find that the counties showing high preparedness levels for the first TC consistently show high preparedness levels for the subsequent one. Power outages induced by the first TC significantly increase preparedness for the subsequent TC (e.g., approximately 13% for mobility needs preparedness and 24% for structural reinforcement preparedness when increasing one-unit customers out). We identified spatial dependency in preparedness across counties. Power outage experiences in first TCs show statistically significant spillover effects on neighboring counties' preparedness levels for second TCs. Throughout sequential TCs, people with access and functional needs consistently show lower preparedness levels.

# Introduction

Driven by rising sea levels and increasing precipitation, destructive hurricanes and tropical storms are becoming increasingly likely to strike coastal areas in quick succession[1]. Sequential tropical cyclones (TCs) — two TCs making landfall in the same region within a short time — can cause compounded damage to affected areas[2]. Severe storms may repeatedly strike coastal communities while they are still suffering from the damages caused by a previous one[1,2]. A recent example is Hurricane Helene (2024) and Hurricane Milton (2024) which sequentially made landfall in Florida. Hurricane Helene struck Florida on September 26 as a Category 4 hurricane with maximum sustained winds of 140 mph, the strongest hurricane on record to impact the Big Bend region. It caused 225 fatalities, making it the deadliest Atlantic hurricane since Hurricane Maria in 2017[3]. Less than two weeks later, on October 9, Hurricane Milton hit Florida as a Category 3 hurricane, inflicting repeated damage to homes, businesses, vehicles, and other infrastructure in coastal areas and claiming 24 lives[3]. The latest updated estimates indicate that the economic losses from this sequential TC event could reach up to $100 billion[4,5].

Preparedness for adverse events is critical for enhancing community resilience and enabling effective adaptation to emerging climate-related risks[6,7]. The National Academies define community resilience as "the capacity of human and infrastructure systems to anticipate, absorb, recover from, or more successfully adapt to actual or potential adverse events," emphasizing the integral role of human systems in building resilience[8]. Communities act as the brain of the city, directing its activities, responding to its needs, and learning from its experience[7], highlighting their adaptive and transformative capabilities in promoting resilience to adverse events[9,10]. Sequential TC hazards can pose additional challenges to population preparedness. Repeated events strain community resilience due to depleted resources and challenge individuals' mental resilience, affecting their capacity and willingness to prepare for, respond to, and recover from consecutive hazards[11]. The probability of sequential TC hazards occurring is projected to increase steadily[1]. Under a high-emissions scenario, the annual occurrence probability of two consecutive category 4 or 5 hurricanes striking the same region could exceed 1% by the end of the century[2]. This emergence of sequential TC hazards highlights the importance of studying and investigating people's situational preparedness dynamics for such hazards.

At the individual or household level, TC preparedness includes short-term (e.g., evacuation, storing food and drinks, and reinforcing structural safety) and long-term measures (e.g., purchasing insurance, stockpiling essentials, and setting mitigation funds)[12–16]. Existing research has identified factors affecting TC preparedness, including socioeconomic status, risk perception, hazard experiences, and access and functional needs (AFN). Findings consistently show that those with lower income, minority racial or ethnic backgrounds, and lower educational attainment are often less proactive in TC preparedness [17,18]. Particularly, individuals with AFNs, including but not limited to those with disabilities, limited language proficiency, restricted mobility, and/or economic advantages as well as children and seniors, face additional challenges in preparing for TCs [19,20]. However, those individuals have received limited attention in previous research and practice [21,22]. Furthermore, studies also recognize that past hurricane experiences and perceived personal risk would increase the likelihood of adopting preparedness measures, such as purchasing home insurance and reinforcing structures[23,24]. However, existing studies do not investigate the temporal dependence of preparedness for sequential TCs, although our previous study on Hurricane Ida (2021) identified a potential association that a previous hurricane would affect the preparedness for the subsequent one[13].

The existing literature has predominantly analyzed preparedness decisions using surveys in the absence of disasters or well after disasters had occurred[25]. Cross-sectional surveys only focus on a specific region and a single event. They often fail to capture local adaptation strategies and policies over time or across different regions[26]. A complementary strand of research utilizes big data to explore the dynamics of preparedness for individual hurricanes. For instance, Dargin et al. (2021) utilized GPS trajectory data to analyze preparedness behaviors for Hurricane Harvey, revealing a statistically significant increase in visits to specific points of interest (POIs) before Harvey, such as gasoline stations, grocery stores, and insurance carriers[27]. Deng et al. (2021) used high-resolution mobility data to explore disparate evacuation patterns among social groups before, during, and after Hurricane Harvey (2017)[12]. Li et al. leveraged SafeGraph mobility data to examine income and racial disparities in preparedness for Hurricane Ida (2021) across multiple affected states[13]. Nevertheless, no existing studies have investigated preparedness dynamics in sequential TC scenarios. To address this gap, we investigate individual- or household-level situational preparedness dynamics for sequential TC hazards using big mobility data.

In this study, we focused on TCs that made landfall with at least tropical storm intensity. Since there is no official definition for a sequential TC event, we focused on cases where two TCs made landfall in the same state within the contiguous United States within 21 days, ensuring the sequential TC events occurred within a typical recovery period for a single TC hazard[28]. We identified six pairs of sequential TCs that affected seven U.S. states (i.e., Louisiana, Texas, New Jersey, New York, North and South Carolina, and Florida) between 2020 and 2024. Using big mobility data, we identified three situational TC preparedness types (i.e., preparedness for mobility needs, daily supplies, and structural reinforcement) by identifying statistically significant increases in visits to POIs (e.g., gasoline stations, grocery stores, and building material dealers). We account for a comprehensive set of factors influencing preparedness behaviors, including forecast TC wind intensity, experienced TC wind intensity and infrastructure disruptions, access and functional needs, and socioeconomic variables. We designed a group of regression analyses to investigate (1) the combined effects of sequential TCs' intensity and landfall sequence on people's preparedness behaviors, (2) the role of individuals' AFN in preparedness disparities, and (3) the impacts of first TCs on the preparedness for the second TCs. This study advances the understanding of the human dimensions of TC events, providing insights into disaster management policies for back-to-back natural hazards.

## Results

**Six pairs of sequential TCs landfalling in the US from 2020 to 2024**

We identified six pairs of sequential TC events between 2020 and 2024 that affected seven states (as shown in **Fig. 1**). These events met the criteria of landfall in the same state as at least a tropical storm characterized by maximum sustained wind speed $\geq 17.5$ m/s (i.e., 34 knots) within 21 days. **Fig. 1** illustrates the forecast track and the forecast uncertainty of the identified sequential TCs.

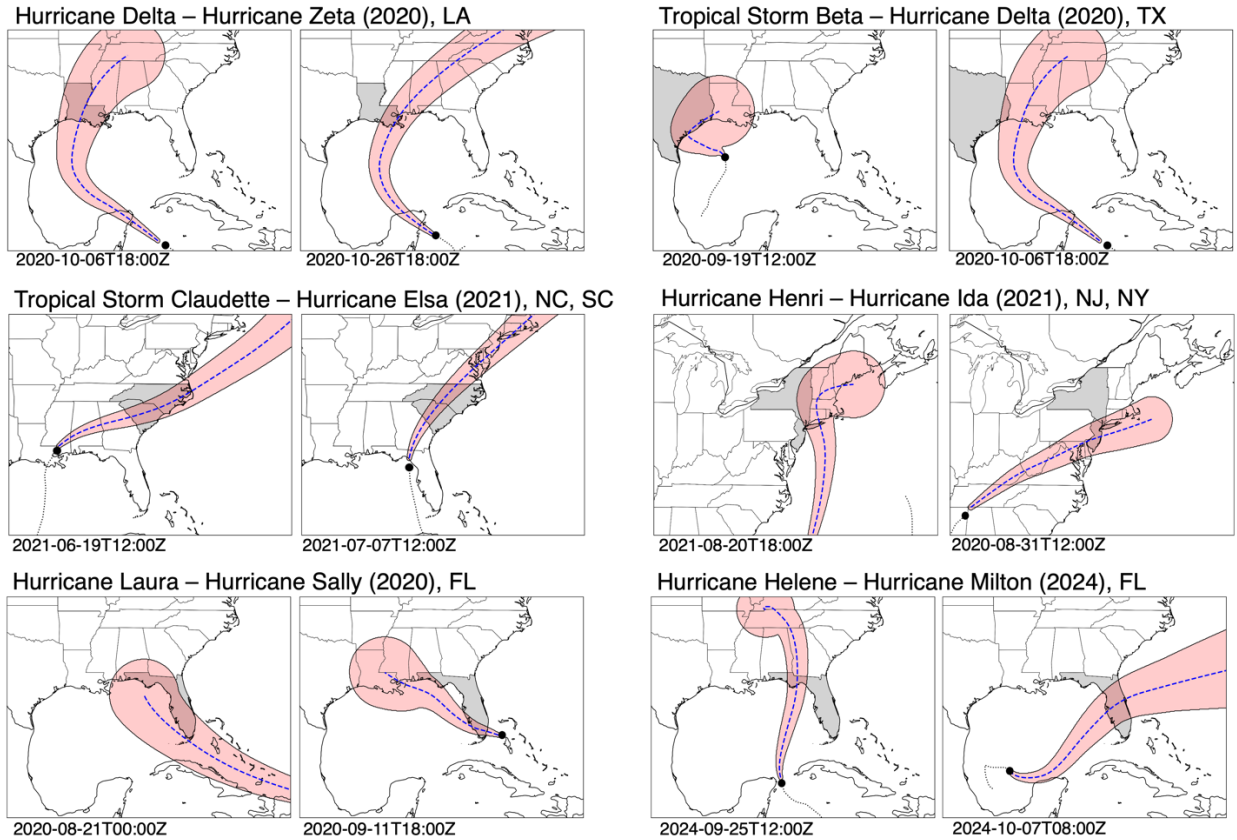

**Fig. 1 Six sequential TC events in the Contiguous United States from 2020 to 2024.** Red areas are cones of uncertainty and dashed lines are the forecast paths of TCs. Data source: National Hurricane Center operational forecast data archive[29].

**Disparate POI visit trends during sequential TC preparedness**

We analyzed people's visits to POIs during sequential TC events. We identified three situational preparedness patterns, characterized by statistically significant increases in POI visits before TC landfall: visiting gasoline stations to prepare for mobility needs, visiting grocery stores to stock up on daily supplies, and visiting building material and supplies dealers to reinforce structural safety, consistent with the findings of our previous work[13,27] and existing literature[30–32]. **Fig. 2** illustrates three types of POI visit trends during Hurricane Helene (2024) and Hurricane Milton (2024), which made sequential landfall in Brevard County, Florida.

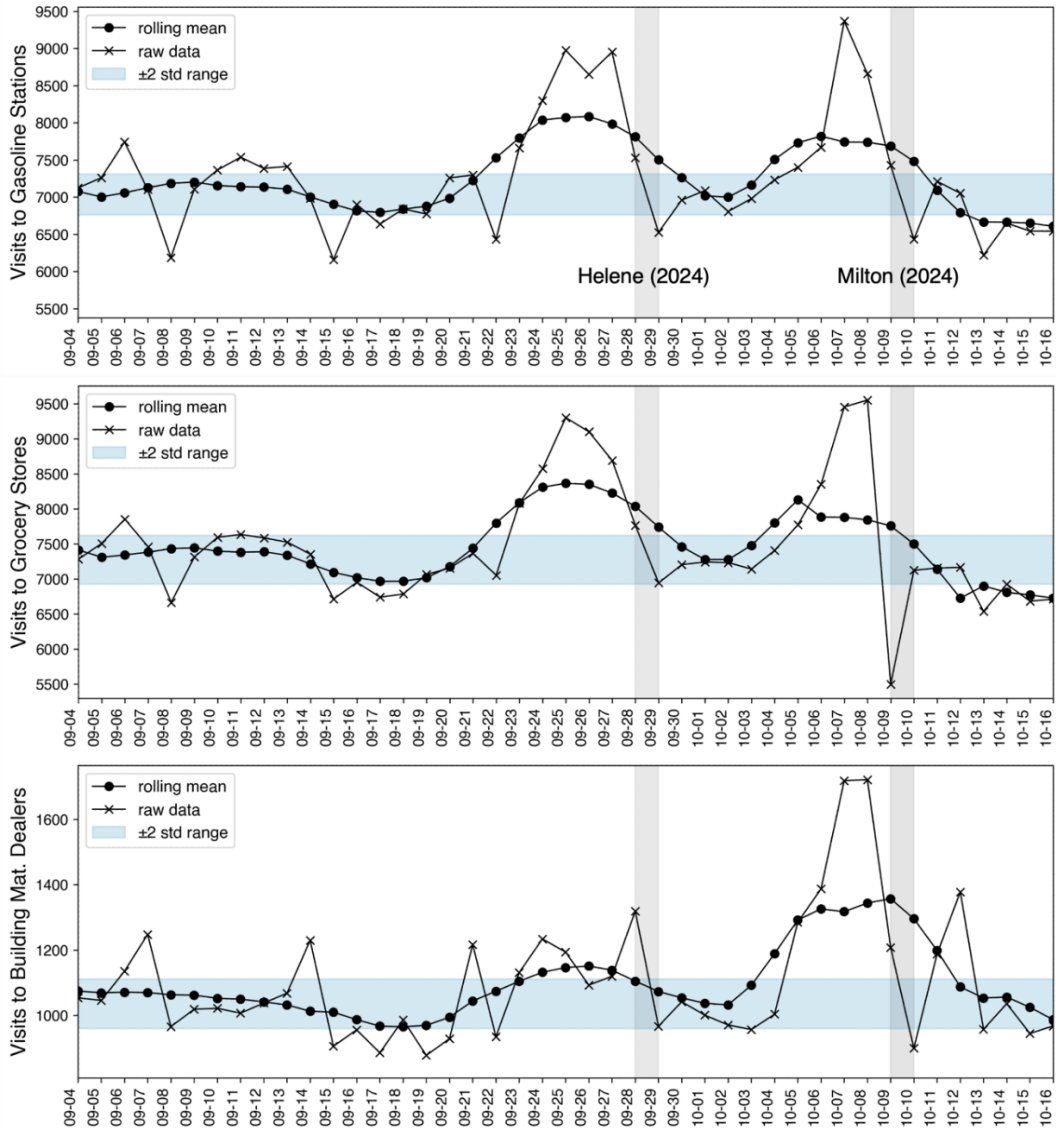

**Fig. 2 Visits to POIs during the Helene-Milton Sequential TC Landfall in Brevard County, Florida.**
We conducted analysis at the county level (i.e., all variables collected or mapped to a county-level resolution) to capture the aggregated individual- or household-level preparedness behaviors. Counties within the same state also experience different intensities of TC due to variations in their distance from the TC path, providing a diverse set of observations. We defined a county-level aggregated preparedness pattern as a statistically significant increase in POI visits before TCs, where visits exceeded two standard deviations above the baseline (see Methods for details). Analyzing the six sequential TCs listed above and the 680 affected county cases, we found disparate situational preparedness patterns across events and illustrated the patterns in **Fig. 3**. For mobility needs preparedness, 165 out of the 680 observations show county-level aggregated preparedness patterns for both TCs, 57 observations show patterns for the first TC only, 101 show patterns for the second TC only, and the remaining 357 show no significant preparedness pattern. For daily supplies (structural reinforcement) preparedness,163 (106) counties have situational preparedness patterns for both TCs, 44 (71) for the first TC only, 146 (92) for the second TC only, and 327 (441) have no preparedness patterns. The level of preparedness efforts, measured by the percentage of increase in POI visits compared to baseline (see methods for details), also varies across TC events. For mobility needs preparedness, 241 of the 680 observations show higher preparedness levels for the first TC than for the second TC, and the other 439 observations show higher preparedness levels for the second TC. For daily supplies preparedness and structural reinforcement preparedness, the comparison is 272 versus 408, and 246 versus 434, respectively. The disparate patterns within and across sequential TCs imply a combined effect of TC intensity and landfall sequence (i.e., the first or second landfalling TC in a sequential TC event) on people's situational preparedness, as investigated below.

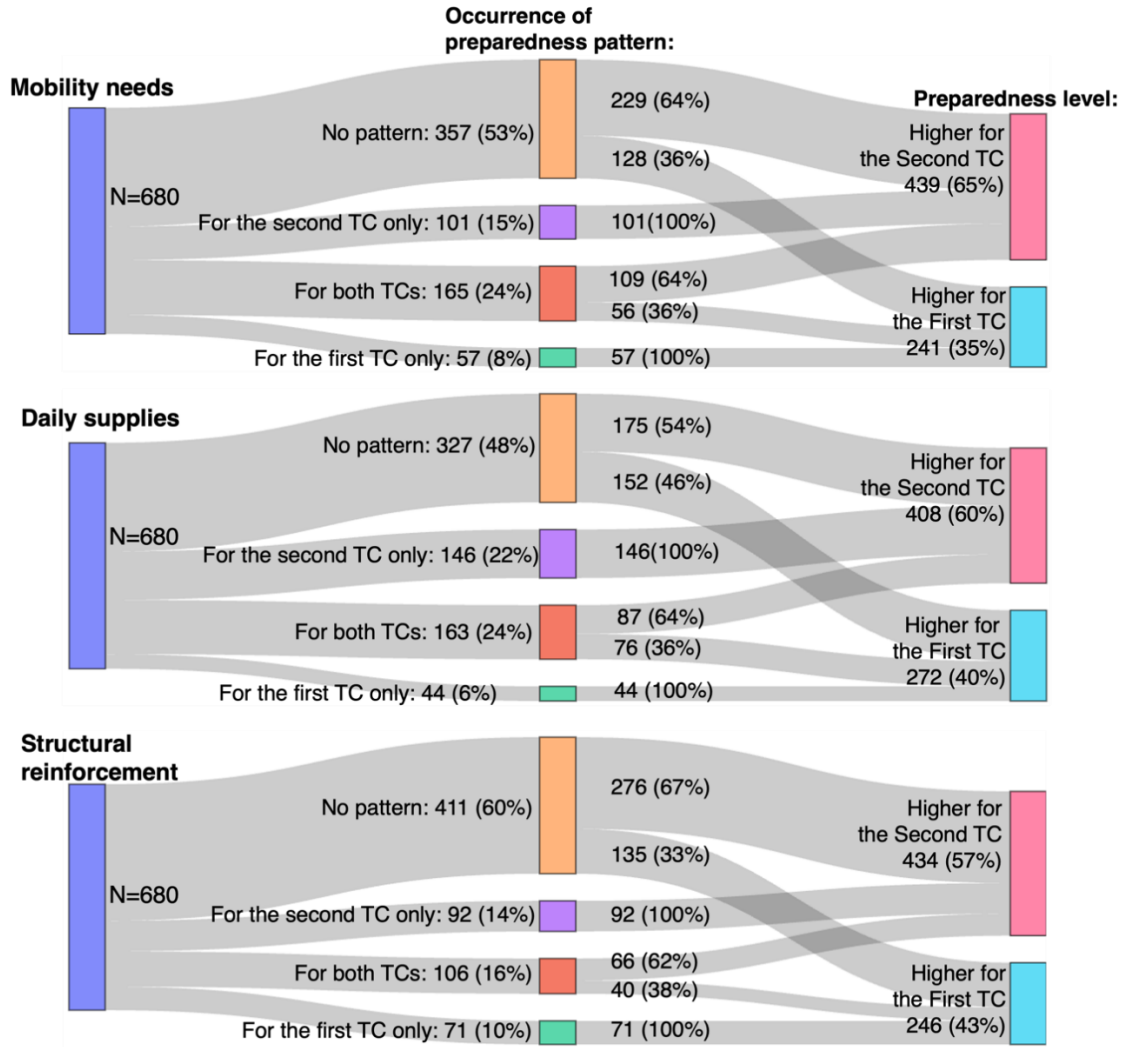

**Fig. 3 Disparate preparedness patterns and preparedness levels for sequential TCs across counties.** We define preparedness level as the percentage of increase in POI visits compared with the baseline and preparedness pattern as the increase in visits exceeding two standard deviations above the baseline. Numbers in the figure are sequential TC cases corresponding to each flow.

**Combined effects of forecast wind and landfall sequence on mobility needs preparedness**

We performed linear spatial regression models (N=1360) to investigate the disparities in the levels of preparedness for the first and second landfall TCs. Each data sample represents a county affected by an individual TC, distinguishing its landfall sequence with a dummy variable, where a subsequent TC is labeled as 1 and the first TC as 0. Linear models capture the continuous changes in the level of preparedness efforts and quantify the variances. We used the level of preparedness efforts, measured by the percentage of a county's increase in POI visits compared to the baseline (see methods for details), as the dependent variable. Table 1 presents the regression results on mobility needs preparedness using Multiple Linear Regression (MLR, the baseline model) and Spatial Durbin Model (SDM). We presented the results of other supplementary models (Spatial Error Model and Spatial Lag Model) in supplementary Table S1. SDM includes controls for the spatial lag of both the dependent and selected independent variables to return an unbiased estimate of the impact of the covariates of interest. Among the models we tested, SDM has the highest log-likelihood value (1342.42), justifying it as the most suitable model to explain the factors influencing preparedness for sequential TCs. In the following sections, we will mainly explain our findings based on SDM (i.e., Model 2).

Table 1 Linear spatial regression models on mobility needs preparedness levels

| | Preparedness level for mobility needs | |
| --- | --- | --- |
| | Model 1(MLR) | Model 2(SDM) |
| Landfall sequence | 0.0314(0.0086)*** | 0.0202(0.0080)* |
| Forecast Wind speed (m/s) | 0.0053(0.0010)*** | 0.0034(0.0009)*** |
| Sequence × Forecast wind speed | -0.0017(0.0006)*** | -0.0012(0.0006)* |
| Prop. Population. Under 5 | -0.0262(0.0023)*** | -0.0245(0.0021)*** |
| Prop. Population. Over 65 | -0.0037(0.0006)*** | -0.0040(0.0006)*** |
| Prop. Disabled Population | -0.1960(0.0963)* | -0.1664(0.0923) |
| Prop. Population. LimEng | -0.1351(0.0854) | -0.3465(0.0905)*** |
| Prop. Household NoVeh | -0.0895(0.0895) | -0.1323(0.0883) |
| Median Income (in 1,000) | -0.0013(0.0003)*** | -0.0012(0.0003)*** |
| Prop. White Population | -0.0163(0.0229) | -0.0413(0.0283) |
| Point of interest density ($m^{-2}$) | 0.0050(0.0110) | -0.0017(0.0145) |
| Spatial lag $\rho$ | | 0.4321(0.0325)*** |
| LM lag | 152.20*** | |
| LM error | 175.80*** | |
| Log likelihood | 1261.67 | 1342.42 |
| Observations | 1360 | 1360 |

*Note:* All regression contains fixed effects for state heterogeneity. See supplementary Table S1 for other state dummy variables and spatial lag variables.
***$p<0.001$, **$p<0.01$, *$p<0.05$.

In model 2, forecast wind speed, landfall sequence, and their interaction term are statistically significantly correlated with the mobility needs preparedness level. Specifically, every unit increase (i.e., 1 m/s) in the forecasted wind speed corresponds to an average increase of 0.34% (95% CI: [0.16%, 0.52%]) in mobility needs preparedness. After controlling for other factors, second landfall TCs receive an average of 2.02% (95% CI: [0.45%, 3.59%]) more preparedness than the first landfall. However, the coefficient of the interaction term indicates that for each unit increase in forecast wind speed, the level of mobility needs preparedness for the first TC is 0.12% (95% CI: [0.01%, 0.23%]) higher than the preparedness for the second one. The marginal effect of forecast wind speed on preparedness level diminishes for the second hurricane relative to the first TC.

The combined effects of landfall sequence and forecast wind speed reveal that populations are more alert to the second landfall TCs but appear less sensitive to increasing forecast wind speed. From a psychological perspective, experiencing back-to-back hurricanes may lead to psychological fatigue, reducing responsiveness to risk signals such as forecast wind speed[33–36]. From a resource perspective, preparedness for the first TCs would consume essential but limited resources (e.g., money and supplies), leaving fewer means for subsequent preparedness for the second TCs[15]. These dynamics highlight that back-to-back hazards would result in compounded risks for communities, and we will further elaborate on it in the Discussion section.

**Spatial dependence and state fixed effects in mobility needs preparedness**

The spatial term in SDM ($\rho = 0.4321$, 95% CI: [0.3685, 0.4958]) indicates the presence of spatial dependence. The spatial terms in the spatial error model and spatial lag model consistently demonstrate the spatial dependence of preparedness between counties. Throughout the linear spatial regression models, we controlled for state-level fixed effects by including dummy variables for each state. We used Florida as the baseline category, so it does not appear as a variable in Model 2. The statistically significant negative coefficients (see supplementary Table S1) of Texas, New York, New Jersey, and South Carolina Indicate that counties In these states have an average of lower preparedness efforts compared to Florida, holding other factors constant. Louisiana and South Carolina show no statistical significance.

**Access and functional needs potentially restrict mobility needs preparedness**

AFN factors in Model 2 exhibit statistically significant associations with the mobility needs preparedness level. The percentage of people under 5 (coefficient -0.0245, 95% CI: [-0.0287,-0.0203]), the percentage of people over 65 (coefficient -0.0040, 95% CI: [-0.0053,-0.0028]), limited language proficiency (coefficient -0.3465, 95% CI: [-0.5239,-0.1690]) are negatively correlated with preparedness level. It suggests that AFNs potentially constrain populations' capability to visit gasoline stations for mobility needs. The proportion of disabled people and the lack of vehicle availability show no statistically significant evidence. We used POI density to estimate the resource availability for preparedness actions, which is also not statistically significant (coefficient-0.0017, 95% CI: [-0.0300,0.0267]). This lack of significance may indicate that simply having a higher density of resources, such as stores and service centers, does not necessarily translate to increased preparedness[37]. Factors such as accessibility, transportation options, and public risk awareness play a more critical role in shaping preparedness behavior.

**Effects of forecast wind speed and landfall sequence on other preparedness types**

The statistical evidence of the combined effects of forecast wind speed and landfall sequence is weaker on preparedness for daily supplies and structural reinforcement. We illustrated the effects of forecast wind speed and landfall sequence in Fig. 4 and presented the full regression analysis in Supplementary Tables S2. The combined effects are not statistically significant for these two preparedness types at the 95% confidence interval. One possible explanation is that the count of visits to grocery stores and building material dealers does not capture the extent of preparedness as effectively as visits to gasoline stations. Factors such as the average one-time payment amount, daily sales, and the types of goods purchased are also meaningful indicators of situational preparedness, but these aspects cannot be reflected solely by visit counts.

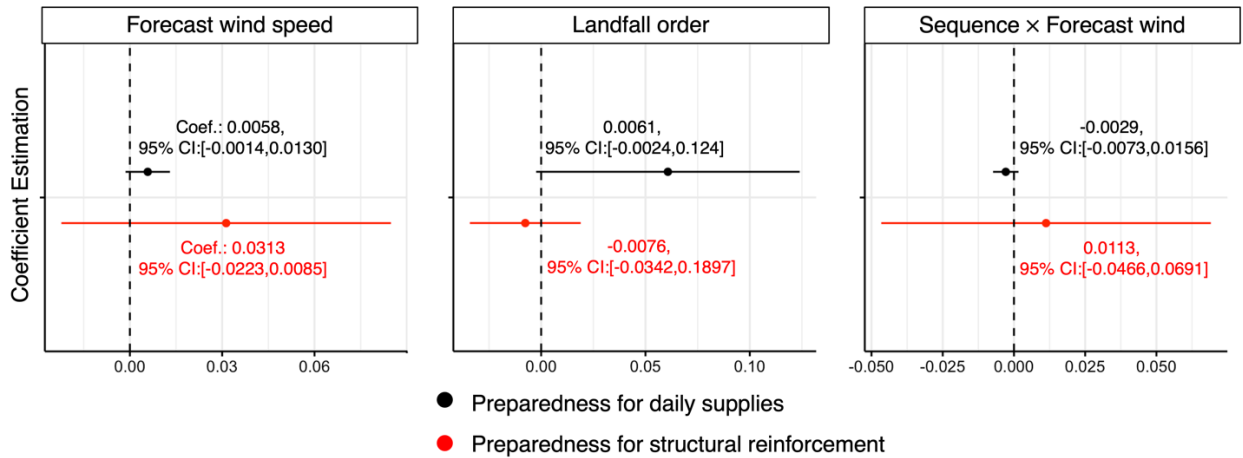

**Fig. 4 Combined effects of forecast wind speed and landfall sequence on daily supplies and structural reinforcement preparedness.** Numbers in the figure represent coefficients of variables, controlling for other variables. The 95% CI represents the confidence interval at a 95% confidence level.

**First landfall TCs influence the preparedness for the subsequent TCs**

We also developed SDMs (N=613) to investigate the first TCs' impacts on the preparedness for a subsequent TC, setting the preparedness level for the second TC as the dependent variable and the preparedness level for the first TC as one of the independent variables. Each data sample describes a county affected by two TCs sequentially. **Table 2** shows the variables and results of the three preparedness types. Notably, the county-level preparedness levels for the two sequential TCs are positively and statistically significantly correlated across three preparedness types, mobility needs preparedness with a coefficient 0.8688 (95% CI: [0.7912, 0.9464], daily supplies preparedness with a coefficient 2.1030 (95% CI: [2.0019, 2.2041], and structural reinforcement preparedness with coefficient 0.9923 (95% CI: [0.9415, 1.0431]. This suggests that counties with high levels of preparedness for the first TCs also tend to have high levels of preparedness for the second TCs, potentially reflecting a learning or habitual effect developed after the first TC. This also reinforces our decision to account for county-level heterogeneity using spatial terms and fixed effects. Additionally, peak power outage from a previous TC potentially increases people's preparedness for the subsequent TC. Every unit increase in previous peak power outage experience corresponds to an average increase of 12.62% (95% CI: [3.3%, 21.9%]) mobility needs preparedness and 25.05% (95% CI: [2.26%, 47.84%]) structural reinforcement preparedness. Maximum experienced wind from previous TC slightly decreases people's mobility needs preparedness for the subsequent TC. Every unit increase in the first TC maximum experienced wind speed corresponds to an average decrease of 0.18% (95% CI: 0.06%,0.30%)). We found statistically significant spillover effects of power outage experiences on the mobility needs preparedness (coefficient 0.2106, 95% CI: [0.0060%, 0.4152]) and daily supplies preparedness (coefficient 1.2568, 95% CI: [0.5610%, 1.9525%]) for the second TCs. The results indicate the spatial interdependency of infrastructures; power outages in neighboring counties will affect the preparedness of the focal county.

**Table 2 SDMs on the second TC preparedness levels**

| | Dependent variable: Second TC preparedness level | | |
|---|---|---|---|
| | Mobility needs | Daily supplies | Struct. reinforcement |
| First TC preparedness level | 0.8707(0.0396)*** | 2.0911(0.0517)*** | 0.9919(0.0260)*** |
| First TC Peak Prop. Power outage | 0.1269(0.0473)** | -0.1226(0.1622) | 0.2424(0.1162)* |
| First TC Power outage duration (h) | -0.0018(0.0015) | 0.0018(0.0050) | 0.0010(0.0037) |
| Experienced first TC max wind (m/s) | -0.0018(0.0006)** | -0.0019(0.0021) | 0.0009(0.0015) |
| First TC max daily rainfall (mm) | -0.0023(0.0579) | 0.0698(0.1938) | -0.0204(0.1423) |
| Second TC forecast wind (m/s) | 0.0014(0.0006)* | 0.0037(0.0021) | 0.0035(0.0015)* |
| First TC Peak Prop. Power outage, Lagged | 0.2106(0.1044)* | 1.2568(0.3550)*** | 0.0731(0.2564) |
| $\rho$ | 0.1142(0.0488)* | -0.0280(0.0420) | -0.0133(0.0405) |
| N | 613 | 613 | 613 |
| Rsq | 0.5747 | 0.7657 | 0.7314 |

*Note:* See supplementary Table S3 for other covariate variables, including state dummy variables and spatial lag variables. We excluded sequential TCs Helene-Milton due to a lack of power outage data in 2024.
***p<0.001, **p<0.01, *p<0.05.

**The county-level aggregated preparedness pattern**

In linear models, we measured the preparedness efforts by the increase in POI visits compared to a baseline. Considering that random fluctuations in POI visits could introduce bias into the assessment of POI visits, noise exists in linear models. To verify our findings, we used the occurrence of aggregated preparedness patterns as a complementary method to evaluate county-level preparedness efforts and demonstrate the robustness of the results. A statistically significant increase in POI visits indicates that the preparedness efforts are sufficiently high to manifest at the county level. These increases surpass normal fluctuations and exceed the baseline mean by more than two standard deviations, thereby highlighting a robust preparedness pattern.

We performed logistic spatial models (N=1360) with binary dependent variables (i.e., the presence or absence of county-level preparedness patterns) while keeping the same independent variables as linear SDM in Table 1. **Fig. 5** confirms the combined effects of TC forecast wind speed and landfall sequence. Consistent with linear models, higher forecast wind speeds are statistically significantly associated with greater mobility needs preparedness (coefficient 0.157, 95% CI: [0.084, 0.230]). Still, its effect diminishes for the second TC compared with the first TC (coefficient -0.068, 95% CI: [-0.110, -0.027]). We find the statistically significant combined effect for daily supplies preparedness, while the results of the linear SDM are insignificant. Given that the two models measure preparedness efforts in different ways, we consider this slight discrepancy acceptable, especially as daily supplies preparedness is close to significance at the 95% confidence level in the linear SDM. Structural reinforcement preparedness has no statistical significance in the combined effects. We consistently found spillover effects in preparedness efforts across counties. We presented the full regression results in supplementary Table S4.

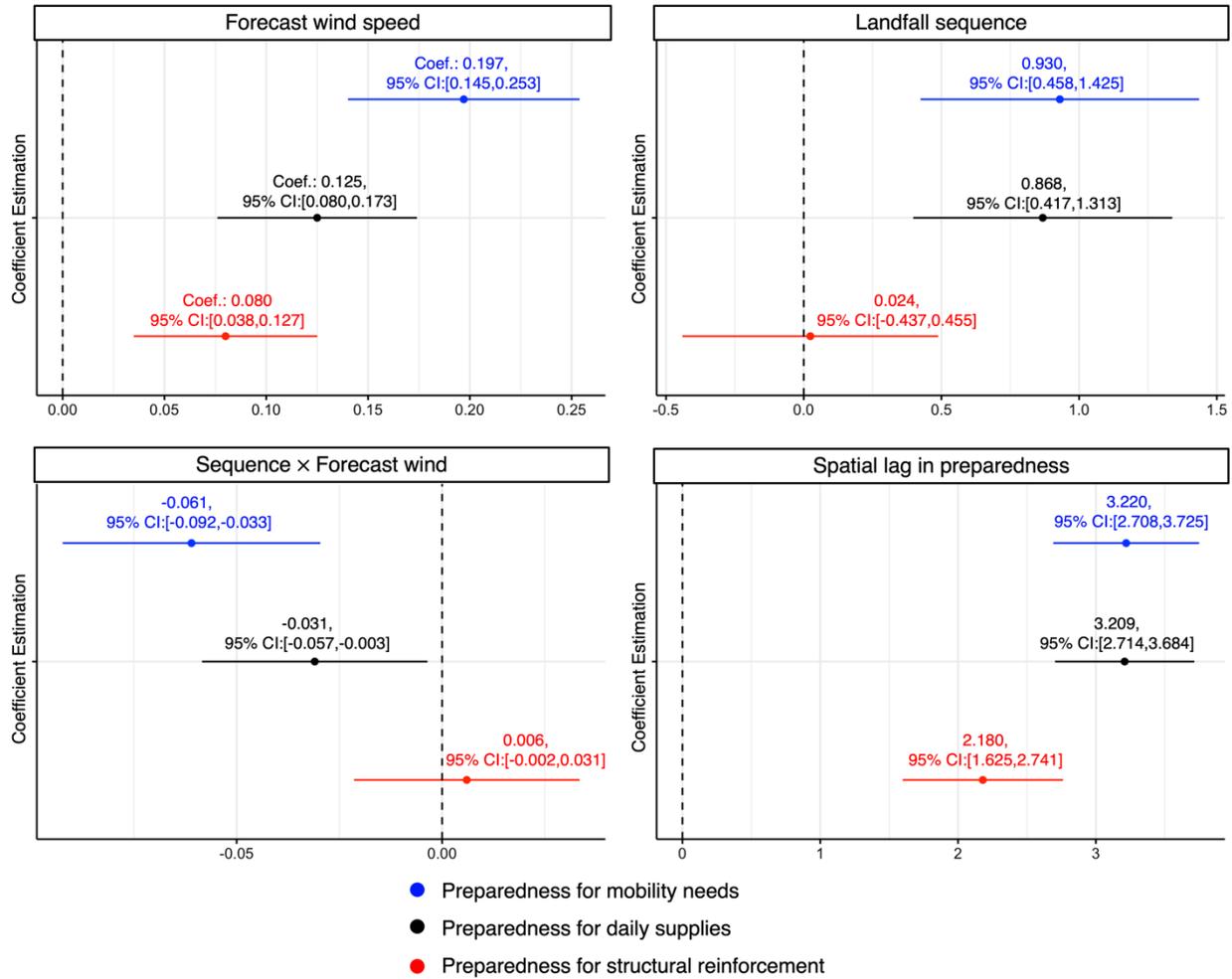

**Fig. 5 Effects of forecast wind speed, landfall sequence, and spatial effects on the county-level aggregated preparedness pattern.** Numbers in the figure represent coefficients of variables, controlling for other variables. The 95% CI represents the confidence interval at a 95% confidence level.

We performed logistic spatial models (N=613) on the second landfall TCs to further verify the first TC's influences on the preparedness for the subsequent one. We used the presence (or absence) of the preparedness pattern as the dependent variable and kept the same independent variables in Table 2. The results show a statistically significant correlation between preparedness for the previous TCs and preparedness for the subsequent ones (illustrated in **Fig. 6**), with a coefficient of 2.530 (95% CI:[1.964, 3.096]) for mobility needs preparedness, 2.084 (95% CI:[1.525, 2.643]) for daily supplies preparedness, and 1.418 (95% CI:[0.918, 1.918]) for mobility needs preparedness. The positive coefficients indicate that counties with higher preparedness levels for the first TC also have higher preparedness levels for the subsequent one. Also, we found statistically significant spatial lag in preparedness for the second TCs, with a coefficient of 1.162, 95% CI: [0.221, 2.103] for mobility needs preparedness and 1.091, 95% CI: [0.134, 2.048] for daily supplies preparedness. The results indicate spatial spillover effects in preparedness efforts across counties—a county tends to have high preparedness levels while neighbors have high preparedness levels. These results are consistent with our findings in Table 2. We presented the full regression results in supplementary Table S5.

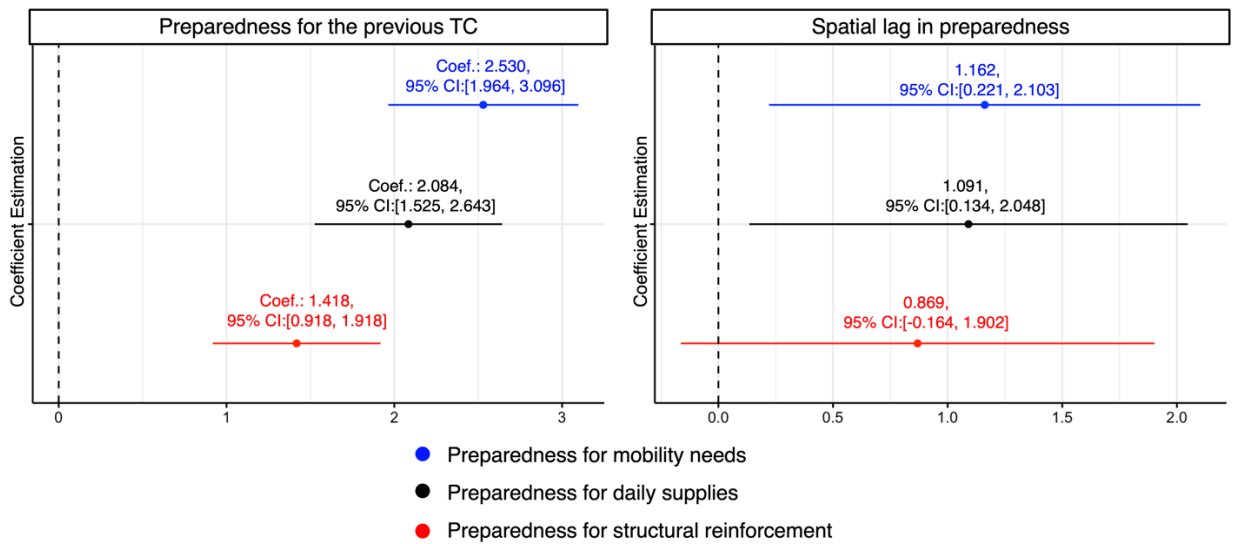

**Fig. 6 Effects of preparedness for the first TC landfall and spatial effects on preparedness for the second TC landfall.** Numbers in the figure represent coefficients of variables, controlling for other variables. The 95% CI represents the confidence interval at a 95% confidence level.

## Discussion

Driven by rising sea levels and increasing precipitation, sequential TC hazards are becoming more likely to happen in the United States. To study situational preparedness dynamics during sequential TC hazards, we investigated three situational preparedness types, namely mobility needs preparedness, daily supplies preparedness, and structural reinforcement preparedness. We identified six sequential TC events that affected Louisiana, Texas, New Jersey, New York, North and South Carolina, and Florida from 2020 to 2024, all landfalling with a higher intensity than a tropical storm (i.e., maximum wind speed over 34 knots). We found that forecast TC intensity and landfall sequence have a combined effect on preparedness. Stronger forecast wind is statistically significantly associated with a higher level of preparedness, indicating people can perceive higher risks from the forecast intensity. Controlling for TC intensity, people tend to have higher levels of preparedness for the second landfalling TC in a sequential TC event. Meanwhile, people are more sensitive to the forecast wind speed for the first TC than the subsequent one. Investigating the first TC's impacts on preparedness for the second TCs, we found that experiencing a severe power outage from a previous TC would increase mobility needs preparedness and structural reinforcement preparedness for the subsequent TC. We did not see this pattern for daily supplies preparedness. We identified spatial dependency in mobility needs preparedness for TCs across counties. Power outage experiences in the first TC show spillover effects on mobility needs and daily supplies preparedness for the second TC.

A higher level of preparedness for the second landfalling TC suggests that the first TC heightened people's risk perception. Our findings support existing research showing that increasing risk perception can promote protective behavior in response to natural hazards, as reflected in a greater willingness to prepare and a stronger public response[38–40]. The consumption of household resources during the first TC may also drive higher preparedness levels for the second hurricane, as people need to replenish essential supplies. Another key finding is that the increasing forecast wind speed has a stronger effect on the increasing preparedness efforts for the first TC than the second TC. It aligns with existing research on the anchoring effects influencing decision-making for future hazards. People may rely on their previous experiences during the first TC to assess the risk of the upcoming second TC[41,42]. Lower sensitivity to the second TC forecast wind speed can also be attributed to mental fatigue[36]. Natural hazards can induce psychological stress and mental health challenge[33–35,43,44]. People might also develop complacency or false sense of security after surviving the first TC. Empirical evidence shows that the government plays a critical role in managing public complacency, especially as an information provider and preparation facilitator[45]. Therefore, it is crucial for local governments to communicate the ongoing risks of sequential storms and emphasize the importance of maintaining sustained preparedness, particularly in coastal regions where the risk of sequential TCs is higher. From another perspective, preparedness efforts for the first TC can deplete essential resources, leaving fewer means for subsequent preparedness. Hence, it is also crucial for the government to strengthen infrastructure and logistics systems to pre-position resources in advance or allocate them during and after sequential TC events. This approach ensures that communities are well-equipped to respond, thereby minimizing the compounded impacts of back-to-back hazards.

Our results suggest that power outage experiences in the first TCs would increase people's preparedness for the subsequent TCs. It aligns with existing research findings that people worrying about power outages tend to have a higher preparedness for potential outages in the future[46]. We observed a weak but statistically significant correlation between peak power outage and experienced wind speed (Spearman correlation coefficient = –0.10, $p < 0.05$), and no statistically significant correlation between experienced wind speed and power outage duration. Power outage risks are not solely driven by wind speed but are closely linked to factors such as grid resilience, wind gusts, vegetation management, and land use[47–52]. We did not control for these factors influencing power outages as they fall outside the scope of this study's focus on preparedness. Power outage experiences during the first TC exhibit spatial spillover effects on neighboring counties' preparedness for the second event, indicating that prior disruptions influence regional responses beyond county boundaries. We also found statistically significant spatial lag in neighboring counties' preparedness levels, suggesting interconnection across neighboring counties' preparedness. These results suggest that county-level preparedness for an approaching TC interacts with neighboring counties due to interdependent infrastructure, sharing ideology, and policy coordination[53–55]. Additionally, state fixed effects reveal variation in preparedness efforts across states, suggesting that differing government ideologies on TC hazards may influence state level TC preparedness[56].

Our analysis reveals statistically significant disparities in preparedness patterns among individuals with Access and Functional Needs (AFNs). Notably, mobility needs are the most severely impacted when TC occurs, showing a strong negative association with the percentage of children under 5, the elderly over 65, individuals with disabilities, households with language barriers, and households without vehicles. These findings underscore the importance of designing targeted, inclusive preparedness strategies that address the specific challenges faced by vulnerable groups, particularly in relation to mobility during disaster events. It is essential to invest in emergency infrastructure that supports accessible transportation options, such as ramps, accessible vehicles, and safe evacuation routes. Additionally, inclusive transportation planning is critical, ensuring that public transit and emergency evacuation systems are designed to accommodate the diverse mobility needs of these populations.

The increasing frequency of sequential TC events presents a new challenge to community resilience. Previous research on compounded hazards mainly focused on simultaneous events, such as compound flooding, heatwaves, and storm-induced power outages [57–59]. Sequential TC hazards strike the same community in succession[1,2]. While a community could be resilient enough to absorb and recover from a single hurricane, the second hurricane might arrive during a vulnerable recovery period, overwhelming the community's resilience capacity. Our study uncovered the disparities between preparedness efforts for the first and second TCs, the first TCs' impacts on preparedness for the second TCs, spatial effects in preparedness across counties, spillover effects of infrastructure disruptions (e.g., power outages), and the role of AFN factors in TC preparedness. Our findings expanded the understanding of people's preparedness for sequential TC hazards, providing insights for improving risk reduction strategies to address the sequentially compound hazards and highlighting the need for targeted outreach and support to assist people with AFNs preparing for sequential TC hazards.

This study has limitations. According to the resilience lifecycle model, an extremely short interval implies that the second TC strikes while the community is still experiencing disruptions and has not yet begun the recovery process[60]. In such cases, situational preparedness may be low due to limited capacity but not a lack of willingness. Conversely, a relatively longer interval allows better conditions for preparedness efforts. We tried incorporating the interval between each pair of sequential TCs in our analysis. Given the availability of mobility data and power outage data, the limited cases provided limited distinct values of intervals for observations, failing to facilitate a robust analysis. We did not find a statistically significant correlation between the interval between two sequential TCs and the level of situational preparedness. Our future works aim to explore additional datasets and develop more robust methods to account for the effects of intervals, distinguishing the impacts of capacity constraints versus willingness on preparedness.

## Methods

**Data pre-processing.** The following sections introduce the data we used. We pre-processed all the data to county resolution according to the TCs affected regions and periods.

**Mobility data.** We used the mobility data provided by Dewey Inc. to assess situational preparedness during sequential TC hazards[61]. The mobility data reports daily visits to POIs and their geographic locations. To account for a consistent weekly of POI visits (i.e. higher visits on weekends than weekdays), we calculated the 7-day rolling average of daily POI visits, representing the preparedness efforts in a continuous 7-day. We defined the preparedness period for a TC as the 7-day before a its landfall. Setting a baseline for each county, we used the average of the 7-day rolling average from four weeks to one week before the sequential TC events (i.e., 21-day before the preparedness period). To estimate the preparedness efforts for each TC event, we calculated the ratio of the maximum 7-day rolling average during the preparedness period (representing peak visits to POIs in a continuous 7-day period) to the baseline value. A ratio value exceeding two standard deviations above the baseline represents a statistically significant increase in POI visits. We defined this threshold exceedance as the presence of a county-level preparedness pattern for the upcoming TC. We defined preparedness level as the percentage of increasing visits to POIs compared to the baseline, which is the ratio value minus 1. The presence (or absence) of a preparedness pattern is the dependent variable in logistic regression models. Preparedness level serves as the dependent variable in linear regression models.

Using the POI locations provided by the mobility data, we calculated POI density for each county by dividing the total number of specific POIs (e.g., gasoline stations) in a county by the county's land area.

**Historical wind.** We utilized the Best Track Data (HURDAT2) from the National Hurricane Center (NHC) to extract the paths and associated wind radii of TCs[29]. HURDAT2 provides data at a 6-hour resolution, which we interpolated to a 1-hour resolution. We used a TC wind field model to produce the TC radial profile of the near-surface rotating wind[62–64]. We recorded the estimated maximum sustained wind speed for each county as the wind impacts from the first TCs.

**Historical rainfall.** We used the NCEP/EMC 4KM Gridded (GRIB) Stage IV Data to estimate historical rainfall during a TC event[65]. We averaged the values of gridded data points within a county's boundary to estimate the rainfall of the county. To assess the rainfall impacts from the first TCs, we calculated the maximum rainfall volume over any continuous 24-hour period[66].

**Forecast wind.** The forecast information was sourced from the NHC Tropical Cyclone Archive, which contains all hurricane forecasts produced since the center's establishment in 1954[29,67]. These forecasts include track and intensity predictions generated every 6 hours throughout the life cycle of a TC. We used the forecast maximum wind speed to estimate the forecast intensity of the TCs.

**Power outage.** The Environment for Analysis of Geo-Located Energy Information (EAGLE-ITM) data, maintained by Oak Ridge National Laboratory, reports the number of customers experiencing electricity outages every 15 minutes at the county level[68,69]. To estimate the total number of customers in a county $c_i$, we used the following formula provided by the data publisher[69]:

$$c_i = p_i * \frac{C}{P}$$

where $C$ is the total number of customers in the state, $P$ is the population of the state, and $p_i$ is the population of county within the state. When $c_{i,out}$ represent the customers without power in county $i$, the extent of power outages, estimated by the power outage rate in a county, follows:

$$Power\ outage\ rate = \frac{c_{i,out}}{c_i}$$

**Factors of sociodemographic and access and functional needs.** To control for the effects of sociodemographic factors on TC preparedness and explore the effects of access and functional needs, we involved the county level household median income, percentage of white population, percentage of people under 5 years old, the percentage of people over 65 years old, the percentage of disabled people aged 5 to 65, the percentage of households with limited English proficiency, and the percentage of households without vehicles in our study. We collected the data mentioned above from US Census Bureau[70].

**Statistical analysis.** We performed spatial regression models to explore: (1) the disparities in preparedness for first TCs and second TCs; (2) the role of AFN in preparedness for TCs; and (3) first TCs' influences on the preparedness for the subsequent TCs. The spatial durbin model (SDM) assumes the spatial dependence exist in both the dependent variable and the independent variables, which follows the baseline specification models:

$$y_{i,j} = \rho W_i y_j + X_{i,j}\beta + \theta W_i Z_{i,j} + \alpha_s + \in$$

where $y_{i,j}$ is county $i$'s preparedness level affected by TC $j$. $W_i$ is the spatial weight matrix associated with county $i$. We used Queen contiguity to define the spatial neighbors of county $i$. $W_i y_j$ thus represent the weighted average of preparedness pattern among county $i$'s neighbors before TC $j$. $X_{i,j}$ is the matrix of independent variables, including climatology, infrastructure, socioeconomic and AFN factors introduced above. $\beta$ is the coefficient associated with $X_{i,j}$. $Z_{i,j}$ is a subset of $X_{i,j}$, including factors affecting infrastructure connection between counties. $\rho$ and $\theta$ characterize the strength of the spatial correlation for the dependent and independent variables, respectively. $\alpha_s$ represents state-level fixed effects. $\epsilon$ is the residual error term. Alternative models include multiple linear regression (MLR), spatial error models (SEM), and spatial lag models (SLR). Each of these alternative approaches serves as a restricted version of the SDM, imposing certain parameter constraints.

We employed spatial logistic models on the presence (or absence) of county-level aggregated preparedness pattern in sequential TC events to demonstrate the robustness of linear models. The spatial logistic model follows a specification baseline :

$$logit\left(P(y_{i,j} = 1)\right) = \rho W_i y_j + X_{i,j}\beta + \theta W_i Z_{i,j} + \alpha_s$$

where $y_{i,j}$ is the vector of binary variables ($y = 1$ for a presence of county-level preparedness pattern and $y = 0$ for an absence). Other terms remain the same with linear models.

## Author contributions

Q.L., F.Z., D.X., and N.L.: Conceptualization; T.L., Q.L., and F.Z.: Methodology; T.L.: Analysis, visualization, writing – Original draft; T.L., Q.L., F.Z., D.X., and N.L.: Writing – Reviewing and Editing

## Competing interest

The authors declare no competing interest.

# Reference

1. Xi, D. & Lin, N. Sequential Landfall of Tropical Cyclones in the United States: From Historical Records to Climate Projections. *Geophys. Res. Lett.* **48**, e2021GL094826 (2021).

2. Xi, D., Lin, N. & Gori, A. Increasing sequential tropical cyclone hazards along the US East and Gulf coasts. *Nat. Clim. Change* **13**, 258–265 (2023).

3. NOAA. U.S. Billion-dollar Weather and Climate Disasters, 1980 - present (NCEI Accession 0209268). NOAA National Centers for Environmental Information https://doi.org/10.25921/STKW-7W73 (2020).

4. FEMA Projects up to $3.5 to $7 Billion in Hurricane Helene Flood Insurance Claim Payments | FEMA.gov. https://www.fema.gov/press-release/20241112/fema-projects-35-7-billion-hurricane-helene-flood-insurance-claim-payments (2024).

5. AccuWeather Report: $500 billion in damage and economic loss estimated after destructive and unprecedented hurricane season. https://www.accuweather.com/en/press/accuweather-report-500-billion-in-damage-and-economic-loss-estimated-after-destructive-and-unprecedented-hurricane-season/1717667.

6. Adger, W. N., Hughes, T. P., Folke, C., Carpenter, S. R. & Rockström, J. Social-Ecological Resilience to Coastal Disasters. *Science* **309**, 1036–1039 (2005).

7. Godschalk, D. R. Urban Hazard Mitigation: Creating Resilient Cities. *Nat. Hazards Rev.* **4**, 136–143 (2003).

8. National Academies. *Disaster Resilience: A National Imperative*. (National Academies Press, 2012).

9. Folke, C. *et al.* Resilience Thinking: Integrating Resilience, Adaptability and Transformability. *Ecol. Soc.* **15**, (2010).

10. Walker, B., Holling, C. S., Carpenter, S. R. & Kinzig, A. P. Resilience, Adaptability and Transformability in Social-ecological Systems. *Ecol. Soc.* **9**, art5 (2004).

11. Kruczkiewicz, A. *et al.* Compound risks and complex emergencies require new approaches to preparedness. *Proc. Natl. Acad. Sci.* **118**, e2106795118 (2021).

12. Deng, H. *et al.* High-resolution human mobility data reveal race and wealth disparities in disaster evacuation patterns. *Humanit. Soc. Sci. Commun.* **8**, 1–8 (2021).

13. Li, Q., Ramaswami, A. & Lin, N. Exploring income and racial inequality in preparedness for Hurricane Ida (2021): insights from digital footprint data. *Environ. Res. Lett.* **18**, 124021 (2023).

14. Wang, D., Davidson, R. A., Trainor, J. E., Nozick, L. K. & Kruse, J. Homeowner purchase of insurance for hurricane-induced wind and flood damage. *Nat. Hazards* **88**, 221–245 (2017).

15. Pan, X., Dresner, M., Mantin, B. & Zhang, J. A. Pre-Hurricane Consumer Stockpiling and Post-Hurricane Product Availability: Empirical Evidence from Natural Experiments. *Prod. Oper. Manag.* **29**, 2350–2380 (2020).

16. Chakravarty, A. K. Humanitarian response to hurricane disasters: Coordinating flood-risk mitigation with fundraising and relief operations. *Nav. Res. Logist. NRL* **65**, 275–288 (2018).

17. Miao, Q. & Zhang, F. Drivers of Household Preparedness for Natural Hazards: The Mediating Role of Perceived Coping Efficacy. *Nat. Hazards Rev.* **24**, 04023010 (2023).

18. Cutter, S. L., Boruff, B. J. & Shirley, W. L. Social vulnerability to environmental hazards. *Soc. Sci. Q.* **84**, 242–261 (2003).

19. Kruger, J., Hinton, C. F., Sinclair, L. B. & Silverman, B. Enhancing individual and community disaster preparedness: Individuals with disabilities and others with access and functional needs. *Disabil. Health J.* **11**, 170–173 (2018).

20. Kailes, J. I. & Enders, A. Moving Beyond "Special Needs": A Function-Based Framework for Emergency Management and Planning. *J. Disabil. Policy Stud.* **17**, 230–237 (2007).

21. Zhang, F. & Xiang, T. Attending to the unattended: Why and how do local governments plan for access and functional needs in climate risk reduction? *Environ. Sci. Policy* **162**, 103892 (2024).

22. Peek, L. & Stough, L. M. Children With Disabilities in the Context of Disaster: A Social Vulnerability Perspective. *Child Dev.* **81**, 1260–1270 (2010).

23. Lindell, M. K. & Hwang, S. N. Households' Perceived Personal Risk and Responses in a Multihazard Environment. *Risk Anal.* **28**, 539–556 (2008).

24. Horney, J. *et al.* Factors Associated with Hurricane Preparedness: Results of a Pre-Hurricane Assessment. *J. Disaster Res.* **3**, 143–149 (2008).

25. Botzen, W. J. W., Mol, J. M., Robinson, P. J. & Czajkowski, J. Drivers of natural disaster risk-reduction actions and their temporal dynamics: Insights from surveys during an imminent hurricane threat and its aftermath. *Risk Anal.* **n/a**,.

26. Usher, K. *et al.* Cross-sectional survey of the disaster preparedness of nurses across the Asia–Pacific region. *Nurs. Health Sci.* **17**, 434–443 (2015).

27. Dargin, J. S., Li, Q., Jawer, G., Xiao, X. & Mostafavi, A. Compound hazards: An examination of how hurricane protective actions could increase transmission risk of COVID-19. *Int. J. Disaster Risk Reduct.* **65**, 102560 (2021).

28. Yuan, F., Esmalian, A., Oztekin, B. & Mostafavi, A. Unveiling spatial patterns of disaster impacts and recovery using credit card transaction fluctuations. *Environ. Plan. B* **49**, 2378–2391 (2022).

29. NHC Data Archive. https://www.nhc.noaa.gov/data/#hurdat.

30. Li, B. & Mostafavi, A. Location intelligence reveals the extent, timing, and spatial variation of hurricane preparedness. *Sci. Rep.* **12**, 16121 (2022).

31. Dooley, D., Catalano, R., Mishra, S. & Serxner, S. Earthquake Preparedness: Predictors in a Community Survey. *J. Appl. Soc. Psychol.* **22**, 451–470 (1992).

32. Bian, R., Smiley, K. T., Parr, S., Shen, J. & Murray-Tuite, P. Analyzing Gas Station Visits during Hurricane Ida: Implications for Future Fuel Supply. *Transp. Res. Rec. J. Transp. Res. Board* **2678**, 706–718 (2024).

33. Kabir, S., Newnham, E., Dewan, A., Islam, M. M. & Hamamura, T. Psychological health declined during the post-monsoon season in communities impacted by sea-level rise in Bangladesh. *Commun. Earth Environ.* **5**, 1–11 (2024).

34. Lowe, S. R., Sampson, L., Gruebner, O. & Galea, S. Psychological Resilience after Hurricane Sandy: The Influence of Individual- and Community-Level Factors on Mental Health after a Large-Scale Natural Disaster. *PLOS ONE* **10**, e0125761 (2015).

35. Weems, C. F. *et al.* The psychosocial impact of Hurricane Katrina: Contextual differences in psychological symptoms, social support, and discrimination. *Behav. Res. Ther.* **45**, 2295–2306 (2007).

36. Davis, C. R., Griffard, M. R., Koo, N. & Pittman, L. R. Resiliency fatigue for rural residents following repeated natural hazard exposure. *Ecol. Soc.* **29**, (2024).

37. Schläpfer, M. *et al.* The universal visitation law of human mobility. *Nature* **593**, 522–527 (2021).

38. Scovell, M., McShane, C., Swinbourne, A. & Smith, D. Rethinking Risk Perception and its Importance for Explaining Natural Hazard Preparedness Behavior. *Risk Anal.* **42**, 450–469 (2022).

39. Bronfman, N. C., Cisternas, P. C., López-Vázquez, E. & Cifuentes, L. A. Trust and risk perception of natural hazards: implications for risk preparedness in Chile. *Nat. Hazards* **81**, 307–327 (2016).

40. Xu, D., Peng, L., Liu, S. & Wang, X. Influences of Risk Perception and Sense of Place on Landslide Disaster Preparedness in Southwestern China. *Int. J. Disaster Risk Sci.* **9**, 167–180 (2018).

41. Paulus, D., de Vries, G., Janssen, M. & Van de Walle, B. The influence of cognitive bias on crisis decision-making: Experimental evidence on the comparison of bias effects between crisis decision-maker groups. *Int. J. Disaster Risk Reduct.* **82**, 103379 (2022).

42. Oki, S. & Nakayachi, K. Paradoxical effects of the record-high tsunamis caused by the 2011 Tohoku earthquake on public judgments of danger. *Int. J. Disaster Risk Reduct.* **2**, 37–45 (2012).

43. Cianconi, P., Betrò, S. & Janiri, L. The Impact of Climate Change on Mental Health: A Systematic Descriptive Review. *Front. Psychiatry* **11**, (2020).

44. Hu, M. D. *et al.* Natural hazards and mental health among US Gulf Coast residents. *J. Expo. Sci. Environ. Epidemiol.* **31**, 842–851 (2021).

45. Wang, X. & Kapucu, N. Public complacency under repeated emergency threats: Some empirical evidence. *J. Public Adm. Res. Theory* **18**, 57–78 (2008).

46. Dominianni, C. *et al.* Power Outage Preparedness and Concern among Vulnerable New York City Residents. *J. Urban Health* **95**, 716–726 (2018).

47. Watson, P. L., Spaulding, A., Koukoula, M. & Anagnostou, E. Improved quantitative prediction of power outages caused by extreme weather events. *Weather Clim. Extrem.* **37**, 100487 (2022).

48. Alemazkoor, N. *et al.* Hurricane-induced power outage risk under climate change is primarily driven by the uncertainty in projections of future hurricane frequency. *Sci. Rep.* **10**, 15270 (2020).

49. Taylor, W. O., Watson, P. L., Cerrai, D. & Anagnostou, E. N. Dynamic modeling of the effects of vegetation management on weather-related power outages. *Electr. Power Syst. Res.* **207**, 107840 (2022).

50. Wanik, D. W., Parent, J. R., Anagnostou, E. N. & Hartman, B. M. Using vegetation management and LiDAR-derived tree height data to improve outage predictions for electric utilities. *Electr. Power Syst. Res.* **146**, 236–245 (2017).

51. Hossain, E., Roy, S., Mohammad, N., Nawar, N. & Dipta, D. R. Metrics and enhancement strategies for grid resilience and reliability during natural disasters. *Appl. Energy* **290**, 116709 (2021).

52. Wang, J. & Lu, F. Modeling the electricity consumption by combining land use types and landscape patterns with nighttime light imagery. *Energy* **234**, 121305 (2021).

53. He, X. & and Cha, E. J. State of the research on disaster risk management of interdependent infrastructure systems for community resilience planning. *Sustain. Resilient Infrastruct.* **7**, 391–420 (2022).

54. Zaman, F. The role of popular discourse about climate change in disaster preparedness: A critical discourse analysis. *Int. J. Disaster Risk Reduct.* **60**, 102270 (2021).

55. Yeo, J., Haupt, B. & Kapucu, N. Alignment between Disaster Policies and Practice: Characteristics of Interorganizational Response Coordination Following 2016 Hurricane Matthew in Florida. *Nat. Hazards Rev.* **22**, 04020052 (2021).

56. Wen, J. & Chang, C.-P. Government ideology and the natural disasters: a global investigation. *Nat. Hazards* **78**, 1481–1490 (2015).

57. Gori, A. & Lin, N. Projecting Compound Flood Hazard Under Climate Change With Physical Models and Joint Probability Methods. *Earths Future* **10**, e2022EF003097 (2022).

58. Feng, D., Shi, X. & Renaud, F. G. Risk assessment for hurricane-induced pluvial flooding in urban areas using a GIS-based multi-criteria approach: A case study of Hurricane Harvey in Houston, USA. *Sci. Total Environ.* **904**, 166891 (2023).

59. Feng, K., Ouyang, M. & Lin, N. Tropical cyclone-blackout-heatwave compound hazard resilience in a changing climate. *Nat. Commun.* **13**, 4421 (2022).

60. CHAPURLAT, V. *et al.* Towards a Model-Based Method for Resilient Critical Infrastructure Engineering How to model Critical Infrastructures and evaluate its Resilience? : How to model Critical Infrastructures and evaluate its Resilience? in *2018 13th Annual Conference on System of Systems Engineering (SoSE)* 561–567 (2018). doi:10.1109/SYSOSE.2018.8428773.

61. Dewey Inc. Dewey | Academic Research Data. https://www.deweydata.io/.

62. Chavas, D. R. Code for tropical cyclone wind profile model of Chavas et al (2015, JAS). (2022) doi:doi:/10.4231/CZ4P-D448.

63. Chavas, D. R., Lin, N. & Emanuel, K. A Model for the Complete Radial Structure of the Tropical Cyclone Wind Field. Part I: Comparison with Observed Structure*. *J. Atmospheric Sci.* **72**, 3647–3662 (2015).

64. Chavas, D. R. & Lin, N. A Model for the Complete Radial Structure of the Tropical Cyclone Wind Field. Part II: Wind Field Variability. *J. Atmospheric Sci.* **73**, 3093–3113 (2016).

65. Du, J. NCEP/EMC 4KM Gridded Data (GRIB) Stage IV Data. Version 1.0. 335570 data files, 2 ancillary/documentation files, 23 GiB UCAR/NCAR - Earth Observing Laboratory https://doi.org/10.5065/D6PG1QDD (2011).

66. Xi, D., Lin, N. & Smith, J. Evaluation of a Physics-Based Tropical Cyclone Rainfall Model for Risk Assessment. (2020) doi:10.1175/JHM-D-20-0035.1.

67. RAL | Tropical Cyclone Guidance Project | Global Repository. https://hurricanes.ral.ucar.edu/repository/.

68. Tansakul, V. *et al.* EAGLE-I Power Outage Data 2014 - 2022. Oak Ridge National Laboratory (ORNL), Oak Ridge, TN (United States). Oak Ridge Leadership Computing Facility (OLCF); Oak Ridge National Laboratory (ORNL), Oak Ridge, TN (United States) https://doi.org/10.13139/ORNLNCCS/1975202 (2023).

69. Brelsford, C. *et al.* A dataset of recorded electricity outages by United States county 2014–2022. *Sci. Data* **11**, 271 (2024).

70. US Census Bureau. Census Datasets. *Census.gov* https://www.census.gov/data/datasets.html.